\documentclass[conference]{IEEEtran}
\usepackage{cite}
\usepackage{amsmath,amssymb,amsfonts}
\usepackage{algorithmic}
\usepackage{graphicx}
\usepackage{textcomp}
\usepackage{xcolor}
\def\BibTeX{{\rm B\kern-.05em{\sc i\kern-.025em b}\kern-.08em
    T\kern-.1667em\lower.7ex\hbox{E}\kern-.125emX}}
\usepackage{subfigure}
\usepackage{algorithm}
\usepackage{tabularx}
\usepackage{booktabs}
\usepackage{xcolor}
\newcommand{\RNum}[1]{\uppercase\expandafter{\romannumeral #1\relax}}

\usepackage{palatino}
\usepackage{tikz}
\usetikzlibrary{shapes.geometric, arrows}

\newtheorem{theorem}{Theorem}[section]

\newtheorem{rem}{Remark}

\usepackage{url}

\usepackage{graphicx}

\begin{document}

\title{Human-machine Hierarchical Networks for Decision Making under Byzantine Attacks}
\author{\IEEEauthorblockN{Chen Quan$^{1}$, Baocheng Geng$^{2}$, Yunghsiang S. Han$^{3}$, Pramod K. Varshney$^{1}$}
\IEEEauthorblockA{
\textit{Email: chquan@syr.edu\quad bgeng@uab.edu\quad yunghsiangh@gmail.com\quad varshney@syr.edu}}
\IEEEauthorblockA{
\textit{$^{1}$Department of Electrical Engineering and Computer
Science, Syracuse University}
}
\IEEEauthorblockA{
\textit{$^{2}$Department of Computer
Science, University of Alabama at Birmingham}
}
\IEEEauthorblockA{
\textit{$^{3}$Shenzhen Institute for Advanced Study, University of Electronic Science and Technology of China}
}
\thanks{Funding from National Key Research and Development Program of China under Grant 2022YFA1004902.}}
\maketitle

\IEEEpeerreviewmaketitle
\begin{abstract}
This paper proposes a belief-updating scheme in a human-machine collaborative decision-making network to combat Byzantine attacks. A hierarchical framework is used to realize the network where local decisions from physical sensors act as reference decisions to improve the quality of human sensor decisions. During the decision-making process, the belief that each physical sensor is malicious is updated. The case when humans have side information available is investigated, and its impact is analyzed. Simulation results substantiate that the proposed scheme can significantly improve the quality of human sensor decisions, even when most physical sensors are malicious. Moreover, the performance of the proposed method does not necessarily depend on the knowledge of the actual fraction of malicious physical sensors. Consequently, the proposed scheme can effectively defend against Byzantine attacks and improve the quality of human sensors' decisions so that the performance of the human-machine collaborative system is enhanced.
\end{abstract}

\section{Introduction}
In high stake scenarios where human lives and assets are at risk, automatic physical sensor-only decision-making may not be sufficient \cite{wimalajeewa2018integrating,sriranga2020human,9443353}. Further, in some circumstances, such as remote sensing and emergency access systems, humans may possess additional side information in addition to the common observations available from both physical sensors and humans. Thus, it may be necessary to incorporate humans in decision-making, intelligence gathering, and decision control. The emerging human-machine inference networks aim to combine humans' cognitive strength and sensors' sensing capabilities to improve system performance and enhance situational awareness.

Unlike physical sensors that can be programmed to operate with fixed parameters, human behavior and decisions are governed by psychological processes which are quite complex and uncertain. Hence, traditional signal processing and fusion schemes can not be adopted directly for integrating sensor measurements with human inputs. It is imperative to construct a framework to capture attributes associated with human-based sources of information so that they can be fused with data from physical sensors.

There have been studies that employ statistical signal processing to address
human-related factors in human-machine collaborative decision making \cite{wimalajeewa2013collaborative,geng2019decision,9443353,wimalajeewa2018integrating,sriranga2020human,8976222,9133140,8969431,gengheteroge,9413745,9747866,geng2021augmented,https://doi.org/10.48550/arxiv.2301.07766,https://doi.org/10.48550/arxiv.2301.07767,https://doi.org/10.48550/arxiv.2301.07789}. For instance, the authors of \cite{wimalajeewa2013collaborative,geng2019decision} studied decision fusion performance when the individual human agents use different thresholds modeled as random variables to make local decisions regarding a given phenomenon of interest (PoI). 
The authors in  \cite{9443353} proposed a hybrid system that consists of multiple human sub-populations, with the thresholds of each sub-population characterized by non-identically distributed random variables and a limited number of machines (physical sensors) whose exact values of thresholds are known. For such a hybrid system, they derived the asymptotic performance at the fusion center in terms of Chernoff information. The authors in  \cite{wimalajeewa2018integrating,sriranga2020human} showed that adding human inputs may or may not improve the overall performance of human-sensor networks, and they derived the conditions under which performance is improved. Furthermore, collaborative decision-making in multi-agent systems was investigated when the rationality of participating humans is modeled using prospect theory \cite{8976222,9133140,8969431,gengheteroge}. 
To a large extent, the literature on human-machine collaborative networks has not considered the distributed nature and the openness of wireless networks in which the physical sensors deployed in the network are low-cost, insecure, and vulnerable to various attacks, e.g., jamming \cite{quan2021strategic}, wiretap, spoofing \cite{jover2014enhancing,gai2017spoofing,ciuonzo2017rician,quan2021strategic}
and Byzantine attacks\cite{zhang2015byzantine,lamport2019byzantine,quan2022enhanced,quan2022efficient,quan2022ordered,quan2022reputation}. In this paper, we are interested in Byzantine attacks, where physical sensors in the network might be compromised and send falsified data to the fusion center (FC).

In contrast to most existing work, this paper aims to construct robust human-machine collaborative decision-making systems. We consider the general scenario where some sensors in the network are compromised by adversaries (Byzantines) so that they send falsified data to human agents. A belief updating and reputation-based scheme, where human agents and physical sensors interact with each other in decision-making, is proposed to mitigate the effect of Byzantine attacks. The proposed scheme consists of three parts: belief updating at human agents, decision-making at the FC, and reputation updating at the FC. In the belief updating part, the human agents make their local decisions based on their observations regarding the PoI and the decisions received from the physical sensors over a short time window. Within this short window, the human agents update their beliefs of the physical sensors' behavioral identities and further update their likelihood ratios (LRs) about the PoI.
The belief updating phase involves collecting information from human agents and physical sensors to contribute to the decision-making at the FC. However, the belief-updating processes at the human agents based on short-term information, i.e., local decisions made by the sensors, may only reflect sensors' behavior over a short period. Consequently, the reputations of physical sensors are also updated over time at the FC to assist in the identification of Byzantine sensors and in mitigating their impact during the decision-making process. Moreover, we study under which conditions human agents can improve the quality of their decisions by using their side information if available. Our simulation results show that the proposed scheme can effectively defend against Byzantine attacks and enhance the quality of human agents' decisions.


\section{SYSTEM MODEL AND THE PROPOSED SCHEME}
In this section, we consider a network model consisting of one FC, $M$ human agents (human sensors), and $N$ physical sensors, all of which make threshold-based binary decisions based on independent observations regarding the PoI. Unlike physical sensors, which employ deterministic thresholds, human sensors are assumed to use random thresholds to make decisions, which account for humans' cognitive biases. We also assume that the human agents have a similar background, e.g., culture, education level, and experience. \footnote{According to studies on human behavior \cite{winquist1998information,kahneman1984choices,9443353,winquist1998information,hodkinson1997careership}, different backgrounds have profound effect on a person's decision-making process, the quality of decisions, as well as the ability to make decisions. To account for the diversity of human populations, we can assume that humans with different backgrounds use random thresholds to follow different distributions. In contrast, random thresholds used by humans with the same background follow the same distribution.} To account for the similar background they are assumed to have, it is reasonable to assume that a known probability distribution characterizes the random thresholds used by human sensors in this work. The thresholds used by the physical sensors are assumed to be the same and deterministic, which are $ \boldsymbol{\tau}=[\tau_1,\dots,\tau_N]^T$. The thresholds used by the human sensors are denoted by $ \boldsymbol{\xi} = [\xi_1,\dots,\xi_{M}]^T$ and they are independent identically distributed (i.i.d.) random variables where $\xi_i$ follows a probability density function (pdf) $f(\xi)$ for $i=1,\dots,M$. In this work, we assume that all the human sensors are honest and put in their best effort to make decisions. We also assume that a fraction $\alpha$ of the $N$ physical sensors are Byzantine nodes and the FC is unaware of the identity of Byzantine nodes in the network. Hence, each sensor has the probability of $\alpha$ being a Byzantine node. As a result of the cognitive biases present in human sensors, some of them might perform worse than others when detecting the PoI. We utilize all useful information from the decisions coming from all the sensors (including physical and human sensors) in the network by employing a human-machine network that is constructed hierarchically, and a belief updating scheme is proposed.

\subsection{Belief-updating scheme}
\begin{figure}[htbp]
\centerline{\includegraphics[width=\linewidth,height=10em]{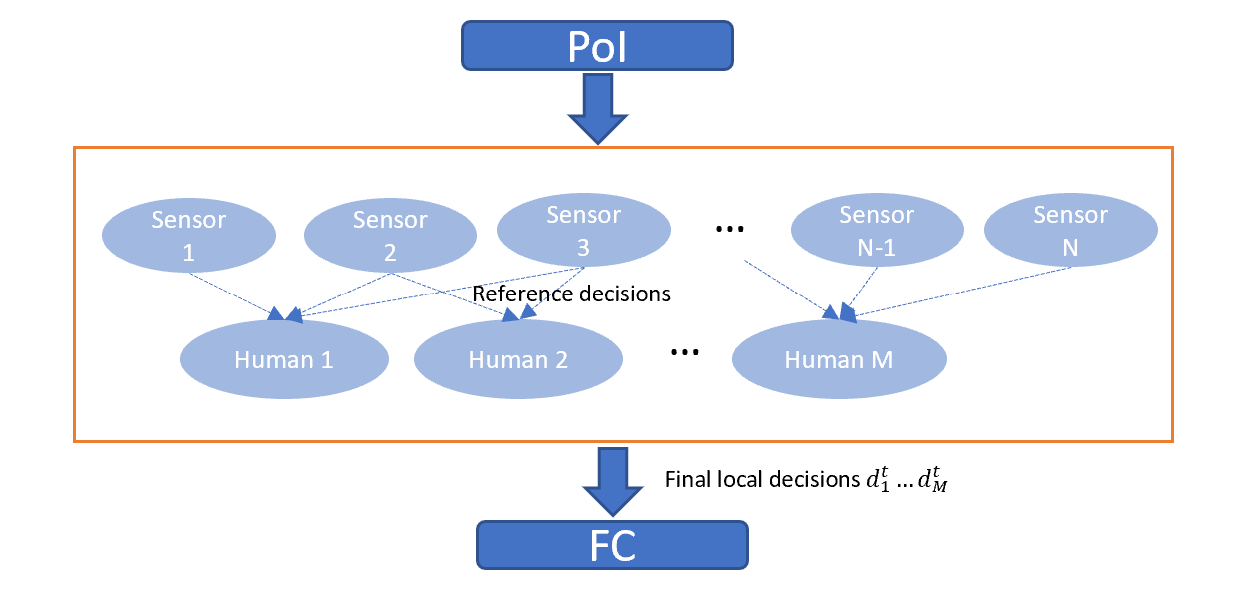}}
\caption{System model}
\label{fig:system}
\end{figure}

The system model is shown in Fig. \ref{fig:system}, where a hierarchical framework is established. All human agents are connected to a small set of physical sensors. Each human agent makes local decisions based on its raw observations and then updates its belief regarding the behavioral identity of the connected physical sensors and its LR based on the local decisions coming from the connected physical sensors' during the time interval $(nT,(n+1)T]$ for $n=0,1,\dots$. 
\begin{rem}
Note that the behavioral identity and the LR are updated at $nT+1$, $nT+2$,\dots until $(n+1)T$. The human sensor uses this updated information to make a decision at time $(n+1)T$. In general, the human sensors collect information from the connected reference physical sensors to update the LR during $(nT,(n+1)T]$. We assume that $T$ is not large so that the true underlying hypothesis does not change during $(nT,(n+1)T]$. At each time step $t=T, 2T, 3T,\dots$, the final decisions made by the FC regarding the presence of the PoI are based on the decisions received from the humans. 
\end{rem}

Let $y_i^t$ and $z_m^t$ denote the observation of sensor $i$ and human $m$ at time $t$, respectively.\footnote{The observations of both human and physical sensors are assumed to be of the same type and are i.i.d..} The LR of physical sensor $i\in\{1,\dots,N\}$ and human agent $m\in\{1,\dots,M\}$ at time $t$ are given as $L_{S,i}^t=\frac{f(y_i^t|\mathcal{H}_1)}{f(y_i^t|\mathcal{H}_0)}$ and $L_{H,m}^t=\frac{f(z_m^t|\mathcal{H}_1)}{f(z_m^t|\mathcal{H}_0)}$, respectively, where hypothesis $\mathcal{H}_1$ indicates the presence of the PoI and hypothesis $\mathcal{H}_0$ indicates the absence of the PoI. Thus, the decision rule of the physical sensor $i$ at time step $t$ is given by
\begin{align}\label{eq:decison_fusion}
       v_i^t=\left\{
\begin{array}{rcl}
1       &      & { L_{S,i}^t\geq \tau_i}\\
0    &      & {otherwise}
\end{array} \right.
\end{align}
Here, we assume that identical physical sensors are deployed, and identical thresholds are utilized at the sensors. Therefore, we have $P_{d,i}= P_{d}$ and $P_{f,i}= P_{f}$ for $i=1,2,\dots,N$. Let $u_{i,m}^{t}$ denote the local decision sent by physical sensor $i$ to the connected human agent $m$ at time $t$, where $m\in\mathcal{M}_m$. If sensor $i$ is malicious, i.e., $i=B$, we assume that $u_{i,m}^{t}=1-v_i^t$; if it is honest, i.e., $i=H$, we assume $u_{i,m}^{t}=v_i^t$. The decision rule of the human agent $m$ based on its raw observation at time step $t$ is given by
\begin{align}\label{eq:decison_fusion_human}
       b_m^t=\left\{
\begin{array}{rcl}
1       &      & { L_{H,m}^t\geq \xi_m}\\
0    &      & {otherwise}
\end{array} \right.
\end{align}

Some key notations used in this paper are listed in Table \ref{table:ref} for the convenience of readers.
\begin{table}
\vspace{-1mm}
\caption{List of Notations Used}
\begin{tabularx}{9.3cm}{@{} rl @{}}
\toprule
  $N$ & number of physical sensors  \\ 
  $M$ & number of human sensors \\
  $\mathcal{M}_m$ & set that consists of physical sensors connected \\
  &to human sensor $m$ \\
  $\mathcal{N}_i$ & set that consists of human sensors connected\\
  &to physical sensor $i$ \\
  $\alpha$ &fraction of Byzantines in the network\\
  $v_i^t$ & the actual local decision made by sensor $i$ at time $t$\\
  $u_{i,m}^{t}$ & the local decision sent by physical sensor $i$ to \\
  &human agent $m$ at time $t$\\
  $b_m^t$ & the local decision made by human sensor $m$ at time\\
  &$t$ which is only based on its raw observations\\
  $d_m^t$ & the local decision made by human sensor $m$ at time\\
  &$t$ when both the local decisions coming from\\
  &connected physical sensors and $b_m^t$ are utilized\\
  $\pi_{h,m,t}$ & probability that $\mathcal{H}_h$ is true at time $t$ for human $m$\\
  $\lambda_{m,t}$ & LR at time $t$ at human sensor $m$\\
  $w_{m,i,t}$ & belief that physical sensor $i$ is honest at human\\
  &sensor $m$ at time $t$\\
  \!\!$w_{m,i,t}(\!u_{i,m}^{t}\!=\!h\!)$ & belief that physical sensor $i$ is honest given $u_{i,m}^{t}\!=\!h$\\
  &at human sensor $m$ at time $t$\\
  $\delta_{m,i,t}$ & LR based on decision coming from physical sensor $i$\\
  & at human sensor $m$ at time $t$\\
  \!\!$\delta_{m,i,t}(\!u_{i,m}^{t}\!=\!h\!)$ & LR given $u_{i,m}^{t}=h$ at human sensor $m$ at time $t$\\
  $r_{m,i,t}$ & probability of $u_{i,m}^{t}=1$ given sensor $i$ is malicious\\
  $q_{m,i,t}$ & probability of $u_{i,m}^{t}=1$ given sensor $i$ is honest\\
  \!$D_{i,H}(\text{or }D_{i,B})$ & probability of $u_{i,m}^{t}=1$ given physical sensor $i$ is\\
  &honest (or malicious) and $\mathcal{H}_1$ is true\\
  \!$F_{i,H}(\text{or }F_{i,B})$ & probability of $u_{i,m}^{t}=1$ given physical sensor $i$ is\\
  &honest (or malicious) and $\mathcal{H}_0$ is true\\
  
  \bottomrule
\end{tabularx}
\label{table:ref}
\end{table}
The belief-updating and decision-making process at each human sensor during any time interval $(nT,(n+1)T]$ proceeds as follows for $n=0,1,\dots$:

\underline{\textbf{Belief-updating}}
\begin{enumerate}
    \item At time $t\in(nT,(n+1)T]$, $q_{m,i,t}$ and $r_{m,i,t}$ are updated, respectively, as
    \begin{align}\label{eq:update_q}
     q_{m,i,t}=&Pr(u_{i,m}^{t}=1|i_{t-1}=H)\notag\\
     =&\pi_{1,m,t-1}D_{i,H}+\pi_{0,m,t-1}F_{i,H}
    \end{align}
    and
    \begin{align}\label{eq:update_r}
    r_{m,i,t}=&Pr(u_{i,m}^{t} =1|i_{t-1}=B)\notag\\
     =&\pi_{1,m,t-1}D_{i,B}+\pi_{0,m,t-1}F_{i,B},
    \end{align}
    for $i\in\mathcal{M}_m$, where $D_{i,X}=Pr(u_{i,m}^{t}=1|\mathcal{H}_1,i=X)=\int_{\tau_i}^{\infty} Pr(y_i^t|\mathcal{H}_1,i=X)\mathrm{d}y_i^t$ and $F_{i,X}=Pr(u_{i,m}^{t}=1|\mathcal{H}_0,i=X)=\int_{\tau_i}^{\infty} Pr(y_i^t|\mathcal{H}_0,i=X)\mathrm{d}y_i^t$ for $X=H$ or $B$. Based on \eqref{eq:update_q} and \eqref{eq:update_r}, the belief that physical sensor $i$ is honest is updated as
    \begin{align}\label{eq:alpha_e1}
        w_{m,i,t}(u_{i,m}^{t}=1)=&\frac{Pr(i_{t-1}=H|u_{i,m}^{t}=1)}{Pr(i_{t-1}=B|u_{i,m}^{t}=1)}\notag\\
        =&\frac{Pr(u_{i,m}^{t}\!=\!1|i_{t-1}\!=\!H)Pr(i_{t-1}\!=\!H)}{Pr(u_{i,m}^{t}\!=\!1|i_{t-1}\!=\!B)Pr(i_{t-1}\!=\!B)}\notag\\
        =&w_{m,i,t-1}\frac{q_{m,i,t-1}}{r_{m,i,t-1}}
    \end{align}
    given $u_{i,m}^{t}=1$ and \begin{align}\label{eq:alpha_e2}
        \!\!w_{m,i,t}(u_{i,m}^{t}\!=\!0)\!=\!&\frac{Pr(i_{t-1}=H|u_{i,m}^{t}=0)}{Pr(i_{t-1}=B|u_{i,m}^{t}=0)}\notag\\
        =&w_{m,i,t-1}\frac{1-q_{m,i,t-1}}{1-r_{m,i,t-1}}
    \end{align}
    given $u_{i,m}^{t}=0$ for $i\in\mathcal{M}_m$, where $Pr(i_{t-1}=B (\text{or }H))$ denotes the probability of sensor $i$ being malicious (or honest) at time step $t$. Note that $Pr(i_{nT+1}=B)=\alpha$ and $Pr(i_{nT+1}=H)=1-\alpha$ for $n=0,1,2\dots$. Hence, the initial belief is $w_{m,i,nT+1}=(1-\alpha)/\alpha$ for $i=0,1,\dots,N$. Given $u_{i,m}^{t}$, the belief that sensor $i$ is honest at time $t$ is $w_{m,i,t}=\frac{Pr(i_{t}=H)}{Pr(i_{t}=B)}=w_{m,i,t}(u_{i,m}^{t}=1)^{u_{i,m}^{t}}w_{m,i,t}(u_{i,m}^{t}=0)^{1-u_{i,m}^{t}}$.
    \item For physical sensor $i$ at time $t$, $\delta_{m,i,t}(u_{i,m}^{t}=h)$ is given by
    \begin{align}
        \!\!\delta_{m,i,t}(u_{i,m}^{t}=1)=&\frac{D_{i,B}\!+\!D_{i,H}w_{m,i,t\!-\!1}}{F_{i,B}\!+\!F_{i,H}w_{m,i,t\!-\!1}}
    \end{align}
    for $h=1$ and 
    \begin{align}
        \!\!\delta_{m,i,t}(u_{i,m}^{t}\!=\!0)=&\frac{(1\!-\!D_{i,B})\!+\!(1-D_{i,H})w_{m,i,t\!-\!1}}{(1-F_{i,B})\!+\!(1\!-\!F_{i,H})w_{m,i,t\!-\!1}}
    \end{align}
    for $h=0$. Hence, given $u_{i,m}^{t}$, the LR at time $t$ is $\delta_{m,i,t}=\delta_{m,i,t}(u_{i,m}^{t}=0)^{1-u_{i,m}^{t}}\delta_{m,i,t}(u_{i,m}^{t}=1)^{u_{i,m}^{t}}$.
    \item \textit{Decision-making at human sensor $m$ at time $t=nT+1$:}
    \begin{equation}
        \lambda_{m,nT+1}\!=\!\frac{\pi_1}{\pi_0}\frac{\beta_m^{b_m^{nT\!+\!1}}(1-\beta_m)^{1-b_m^{nT\!+\!1}}}{\gamma_m^{b_m^{nT\!+\!1}}(1\!-\!\gamma_m)^{1\!-\!b_m^{nT+1}}} \overset{d_m^{nT\!+\!1}=1}{\underset{d_m^{nT\!+\!1}\!=\!0}{\gtrless}}\kappa',
    \end{equation}
    where ${\beta}_m= \int_{\xi_m}^{\infty}f(z_m^t|\mathcal{H}_1)\mathrm{d}z_m^t$, ${\gamma}_m= \int_{\xi_m}^{\infty}f(z_m^t|\mathcal{H}_0)\mathrm{d}z_m^t$ are the probabilities of detection and false alarm for human agent $m$.
    
    \textit{Decision-making at human sensor $m$ at time $t\in{[nT+2,(n+1)T]}$:}
    \begin{equation}
        \!\!\!\!\!\!\lambda_{m,t}\!=\!\lambda_{m,t\!-\!1}\frac{\beta_m^{b_m^{t}}(1\!-\!\beta_m)^{1\!-\!b_m^{t}}}{\gamma_m^{b_m^{t}}(1\!-\!\gamma_m)^{1\!-\!b_m^{t}}}\!\!\prod_{j\in\mathcal{M}_m}\delta_{m,j,t} \overset{d_m^t\!=\!1}{\underset{d_m^t\!=\!0}{\gtrless}}\kappa'
    \end{equation}
\end{enumerate}
\underline{\textbf{Decision-making at FC}}
At the time $t=(n+1)T$, the fusion rule at the FC is given as
\begin{equation}
    \sum_{m=1}^Md_m^{(n+1)T}\overset{\mathcal{H}_1}{\underset{\mathcal{H}_0}{\gtrless}}\kappa,
\end{equation}
where $\kappa$ is the threshold used by the FC.
\underline{\textbf{Reputation-updating at FC}}
At time $t=T,2T\dots$, the reputation of sensor $i$ is given as $r_{i}^t=r_{i}^{t-T}+A_{i}^{t-T}$,
where
\begin{align}\label{eq:update_rule}
    A_{i}^{t-T}=\left\{
    \begin{array}{rcl}
\Delta\frac{c_{i,t}}{|\mathcal{N}_i|}       &      & {c_{i,t}>c_{i,t}/2} \\
-\Delta(1-\frac{c_{i,t}}{|\mathcal{N}_i|})   &      & {otherwise}.
\end{array} \right.
\end{align} 
$|\mathcal{N}_i|$ is the cardinality of $\mathcal{N}_i$, $c_{i,t}\!\!=\!\!\sum_{m\in\mathcal{N}_i}\!\!I(w_{m,i,t})$ where
\begin{align}
       I(w_{m,i,t})=\left\{
\begin{array}{rcl}
1       &      & {w_{m,i,t}>1} \\
-1    &      & {otherwise}
\end{array} \right.
\end{align}
and $\Delta$ is the step size to update the reputation of each physical sensor. According to \eqref{eq:update_rule}, a sensor's reputation increases if most human agents believe it is honest, and vice versa. The more human agents vote in favor of the same decision, the greater the increment in the reputation of sensors.
When $r_{i}^t$ is smaller than a threshold $\eta$, sensor $i$ is identified as Byzantine and initial reputation is $r_{i}^{0}=1$ for $i=0,1,\dots,N$.

\subsection{Human sensors with side information}
Thus far, we have assumed that human and physical sensors only receive i.i.d. observations. In this subsection, we assume that human sensors may also possess side information\footnote{The side information refers to the additional information owned by human sensors which could come from previous professional experience or other sources.} about the PoI other than the common features, which both physical and human sensors can observe.
Assume that the human sensor $m$ possesses the side information $w_m^t$ related to the PoI in addition to the common attribute $z_m^t$ for $m=1,2,\dots,M$. To emulate the actions humans take to incorporate the data gathered from side information and observations into their decision-making process, two operations are employed in this work which are OR operation and AND operation\cite{9443353}.

\paragraph{OR operation}
The decision rule when using the OR operation to include side information is given by
\begin{align}
       e_m^t=\left\{
\begin{array}{rcl}
1       &      & {b_m^t =1 \quad \text{or}\quad w_m^t = 1}\\
0    &      & {otherwise}
\end{array} \right.
\end{align}
where $w_m^t$ is the side information indicating whether $\mathcal{H}_1$ is present or not and is assumed to be binary for human sensor $m$. The accuracy of side information is denoted as $Pr(w_m^t=1|\mathcal{H}_1)=\beta_{m,side}$,  and $Pr(w_m^t=1|\mathcal{H}_0)=\gamma_{m,side}$. We assume that the side information $\{w_m^t\}_{m=1}^M$ is independent among different human sensors.
Given the side information, the likelihoods of $e_m^t$ given $\mathcal{H}_1$ and $\mathcal{H}_0$ are shown, respectively, as\cite{9443353}
\begin{equation}\label{eq_likelyhood_f1_side}
    f(e_m^t|\mathcal{H}_1) \!=\! \beta_{m,side} e_m^t \!+\!(1\!-\!\beta_{m,side})(\bar{\beta}^{e_{m}^t}(1\!-\!\bar{\beta})^{1\!-\!e_{m}^t})
\end{equation}
\begin{equation}\label{eq_likelyhood_f0_side}
    f(e_m^t|\mathcal{H}_0) \!= \!\gamma_{m,side} e_m^t \!+\!(1\!-\!\gamma_{m,side})(\bar{\gamma}^{e_{m}^t}(1\!-\!\bar{\gamma})^{1\!-\!e_{m}^t}),
\end{equation}
where $\bar{\beta}= \int_{-\infty}^{\infty}f(\xi) Pr(b^t=1|\mathcal{H}_1,\xi)\mathrm{d}\xi$ and $\bar{\gamma}= \int_{-\infty}^{\infty}f(\xi) Pr(b^t=1|\mathcal{H}_0,\xi)\mathrm{d}\xi$ are the averaged probabilities of detection and false alarm for all human agents, respectively. Based on \eqref{eq_likelyhood_f1_side} and \eqref{eq_likelyhood_f0_side}, we can derive the probability of detection $P_{d,m,side}^{OR}$ and probability of false alarm $P_{f,m,side}^{OR}$ for human sensor $m$ that adopts OR operation, which are given, respectively, by $P_{d,m,side}^{OR}=\beta_{m,side}+(1-\beta_{m,side})\bar{\beta}$
and $P_{f,m,side}^{OR}=\gamma_{m,side}+(1-\gamma_{m,side})\bar{\gamma}$.
Thus, the overall probability of error for human sensor $m$ using OR operation becomes
\begin{equation}\label{eq:ERR_SIDE_OR}
    P_{e,m,side}^{OR}=\pi_0P_{f,m,side}^{OR}+\pi_1(1-P_{d,m,side}^{OR})
\end{equation}
\paragraph{AND operation}
The decision rule when employing the AND operation to include the human observation and side information is expressed as
\begin{align}
       e_m^t=\left\{
\begin{array}{rcl}
1       &      & {b_m^t =1 \text{ and}\quad w_m^t = 1}\\
0    &      & {otherwise}
\end{array} \right.
\end{align}\label{eq:decision_rule_AND}
Given the side information, the likelihoods of $e_m^t$ given $\mathcal{H}_1$ and $\mathcal{H}_0$ are expressed, respectively, as\cite{9443353}
\begin{equation}\label{eq:likelyhood_f1_AND}
    f(e^t_m|\mathcal{H}_1) = \beta_{m,side} \bar{\beta}^{e_m^t}(1-\bar{\beta})^{1-e_m^t} +(1-\beta_{m,side})(1-e_{m}^t)
\end{equation}
\begin{equation}\label{eq:likelyhood_f0_AND}
    f(e^t_m|\mathcal{H}_0) = \gamma_{m,side} \bar{\gamma}^{e_m^t}(1-\bar{\gamma})^{1-e_m^t} +(1-\gamma_{m,side})(1-e_{m}^t)
\end{equation}

Based on \eqref{eq:likelyhood_f1_AND} and \eqref{eq:likelyhood_f0_AND}, we can derive the probability of detection $P_{d,m,side}^{AND}$ and the probability of false alarm $P_{f,m,side}^{AND}$ for human sensor $m$ that adopts AND operation, which are given, respectively, by $P_{d,m,side}^{AND}=\beta_{m,side}\bar{\beta}$
and $P_{f,m,side}^{AND}=\gamma_{m,side}\bar{\gamma}$.
Thus, the overall error probability for human sensor $m$ using AND operation becomes
\begin{equation}\label{eq:ERR_SIDE_AND}
    P_{e,m,side}^{AND}=\pi_0P_{f,m,side}^{AND}+\pi_1(1-P_{d,m,side}^{AND})
\end{equation}
\paragraph{\underline{Comparison between OR and AND operation}} 
As a result of the above analysis, the performance of each human sensor in terms of error probability is obtained when using the OR operation and when using the AND operation, i.e., $P_{e,m,side}^{OR}$ and $P_{e,m,side}^{AND}$. It is also easy to obtain the averaged error probability of each human sensor without obtaining any side information. It is given by
\begin{align}\label{eq:ERR_NO_SIDE}
    P_{e}=&\pi_0\bar{\gamma}+\pi_1(1-\bar{\beta}),
\end{align}
where $\pi_h$ denotes the probability that $\mathcal{H}_h$ is true for $h=0,1$.
Based on \eqref{eq:ERR_NO_SIDE}, \eqref{eq:ERR_SIDE_AND} and \eqref{eq:ERR_SIDE_OR}, we can derive the conditions under which the quality of human sensors' decisions is improved by utilizing different operations to utilize side information. The derived conditions are stated in Theorem \ref{them:comapre_err}.

\begin{theorem}\label{them:comapre_err}
When the following conditions are satisfied, the side information could help individual human sensors make better decisions. For a specific human sensor $m\in\{1,\dots,M\}$, we have
\begin{itemize}
    \item the quality of decisions is improved by utilizing AND operation when $\frac{\bar{\beta}}{\bar{\gamma}}\leq\frac{\pi_0(1-\gamma_{m,side})}{\pi_1(1-\beta_{m,side})}$.
    \item the quality of decisions is improved by utilizing OR operation when $\frac{\pi_0(1-\bar{\gamma})}{\pi_1(1-\bar{\beta})}\leq\frac{\beta_{m,side}}{\gamma_{m,side}}$.
    \item OR operation performs better than AND operation when $\pi_0(1\!-\!2\bar{\gamma})\gamma_{m,side}\!-\!\pi_1(1\!-\!2\bar{\beta})\beta_{m,side}\leq\pi_1\bar{\beta}-\pi_0\bar{\gamma}$.
\end{itemize}
\end{theorem}
\begin{IEEEproof}
The above conditions can be derived by comparing the value of $P_{e,m,side}^{AND}$ and $ P_{e}$, the value of $ P_{e,m,side}^{OR}$ and $ P_{e}$, and the value of $P_{e,m,side}^{AND}$ and $ P_{e,m,side}^{OR}$.
\end{IEEEproof}

\section{Numerical Results}
Some numerical results are presented in this section. Assume $y_i^t|\mathcal{H}_h\sim \mathcal{N}(\mu,\sigma_h^2)$ and $z_m^t|\mathcal{H}_h\sim \mathcal{N}(\mu_h,\sigma_h^2)$ for $h=0,1$, where $\mu_1=4$, $\mu_0=0$ and $\sigma_1^2=\sigma_0^2=2$. The human thresholds are assumed to follow the Gaussian distribution with parameters $(\mu_{\tau},\sigma_{\tau})$, where $\mu_{\tau}=2$ and $\sigma_{\tau}=2$. We set $N=60$, $M=20$, $T=10$, $\Delta=0.03$, $\kappa'=1$, $\kappa=M/2$, $\eta=0.2$, $\tau_{i}=2$ for $i=1,\dots,N$ and $\pi_1^{nT+1}=\pi_0^{nT+1}=0.5$ for $n=0,1,\dots$. Note that the term 'iterations' used in the following figures refers to iterations during the belief updating phase.

\begin{figure}[htb]
  \centering \includegraphics[width=\linewidth,height=9em]{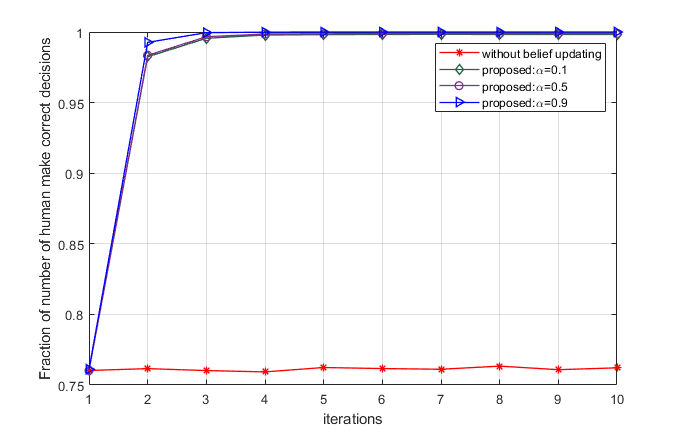}
  \caption{Fraction of number of humans that make correct decisions versus the number of iterations when the system is aware of $\alpha$}
  \label{fig:1_3}
\end{figure}

In Table \ref{table:t1}, we show the comparison of the error probabilities of the systems that adopt CV (Chair-Varshney rule), MR (Majority rule), and MRH (Majority rule with human sensors) when $\alpha$ is known. MRH only utilizes the decisions from human sensors, while MR and CV utilize both human and physical sensors. The MR system uses decisions from all the sensors (including physical and human sensors) by performing a simple majority vote to make a final decision. In contrast, our proposed scheme employs a hierarchical framework to construct the human-machine collaborative network. As seen in Table \ref{table:t1}, MR breaks down when most sensors participating in the decision-making process are malicious. However, our proposed scheme can still achieve comparable performance to the optimal CV rule. We can see in Fig. \ref{fig:1_3} that the fraction of humans making correct decisions increases significantly within a small number of iterations in our proposed scheme, which indicates a rapid improvement in the quality of humans' decisions.
\vspace{-2mm}
\begin{table}[htbp]
\caption{System error probability as a function of $\alpha$}
\label{table:t1}
\centering  
\begin{tabular}{|c|c|c|c|}
\hline
 & $\alpha=0.1$ & $\alpha=0.5$ & $\alpha=0.9$\\
\hline
CV & 2.3e-7 &3.1e-7&4e-7\\
\hline
MR & 7e-5&0.07& 0.996\\
\hline
MRH & 2.5e-3&2.5e-3&2.5e-3 \\
\hline
Proposed & 2.7e-7 &3.7e-7&4.3e-7\\
\hline
\end{tabular}
\end{table}
\vspace{-2mm}

\begin{figure}[htb]
  \centering   \includegraphics[width=\linewidth,height=10em]{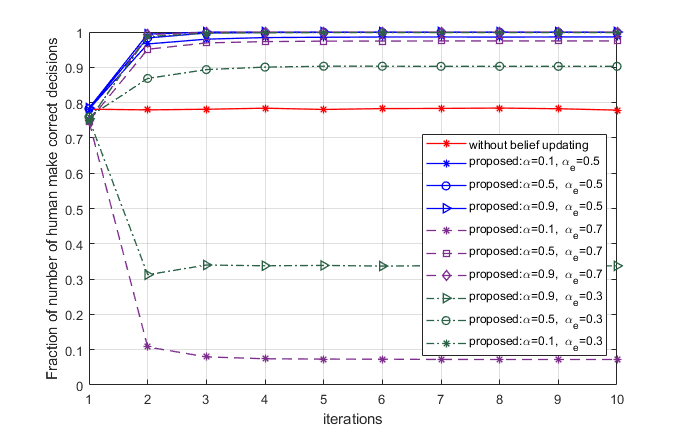}
  \caption{Fraction of the number of humans that make correct decisions versus the number of iterations when the system does not know $\alpha$.}
  \label{fig:4_5}
\end{figure}

\begin{figure}[htbp]
\centerline{\includegraphics[width=\linewidth,height=9em]{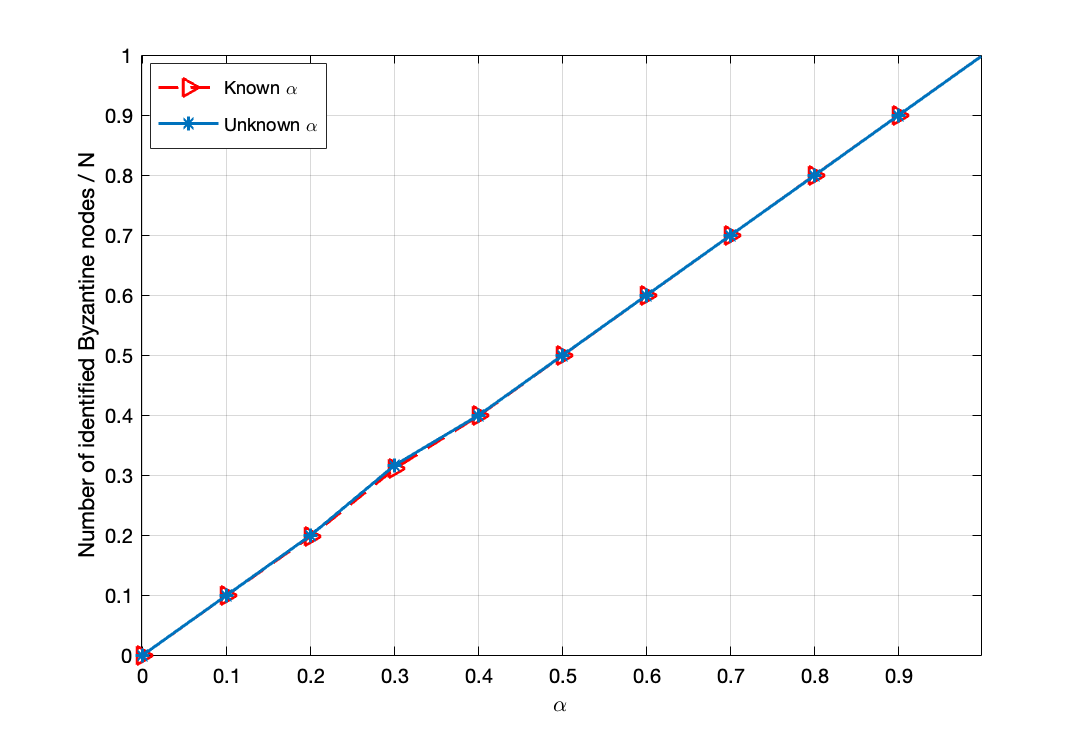}}
\caption{The ratio of identified Byzantine nodes to the total number of sensors versus $\alpha$ for the proposed scheme when $\alpha$ is known and unknown. If $\alpha$ is unknown, we set $\alpha_e=0.5$.}
\label{fig:6}
\end{figure}

Although the proposed scheme requires the knowledge of $\alpha$ when each human sensor updates the belief regarding the behavioral identity of the corresponding physical sensors, this knowledge is not necessarily needed to guarantee a good performance. Choosing an appropriate predefined $\alpha$ can alleviate the performance degradation caused by the absence of knowledge of $\alpha$. In Fig. \ref{fig:4_5} and Table \ref{table:t2}, we compare the performance of the different systems when $\alpha$ is replaced with a predefined value $\alpha_e$ in \eqref{eq:alpha_e1} and \eqref{eq:alpha_e2}. Fig. \ref{fig:4_5} shows the fraction of the number of humans that make correct decisions given different values of $\alpha_e$. We can observe that the system with $\alpha=0.9$ performs worse when $\alpha_e=0.3$ and the system with $\alpha=0.1$ performs worse when $\alpha_e=0.7$. However, $\alpha_e=0.5$ works well for the system with any fraction of Byzantine nodes. This is because when $\alpha_e\gg\alpha$ (or $\alpha_e\ll\alpha$), $\alpha_e$ significantly overestimates (or underestimates) $\alpha$ which results in performance degradation. Thus, $\alpha_e=0.5$ is a good choice when we do not know $\alpha$. Table \ref{table:t2} show that the system that adopts the proposed scheme can outperform the systems that adopt CV, MR, and MRH when $\alpha$ is unknown. Although there is a performance degradation compared to the systems that are aware of $\alpha$, i.e., the performance shown in Fig. \ref{fig:1_3} and Table \ref{table:t1}, the performance degradation is negligible for the proposed scheme. Thus, whether we know the actual $\alpha$ or not, the proposed scheme can always achieve a good performance. In Fig. \ref{fig:6}, we show that the proposed scheme performs well in identifying Byzantine nodes in both cases, i.e., the system is aware/unaware of $\alpha$.



\vspace{-2mm}
\begin{table}[htbp]
\caption{System error probability as a function of $\alpha$ given $\alpha_e=0.5$}
\label{table:t2}
\centering  
\begin{tabular}{|c|c|c|c|}
\hline
 & $\alpha=0.1$ & $\alpha=0.5$ & $\alpha=0.9$\\
\hline
CV & 8e-4 & 1.3e-3 &1.8e-3\\
\hline
MR & 7e-5&0.07& 0.996\\
\hline
MRH & 2.5e-3&2.5e-3&2.5e-3 \\
\hline
Proposed & 1.7e-5 &4.1e-7&3.3e-5\\
\hline
\end{tabular}
\end{table}
\vspace{-2mm}

\begin{figure}[htbp]
\centerline{\includegraphics[width=\linewidth,height=9em]{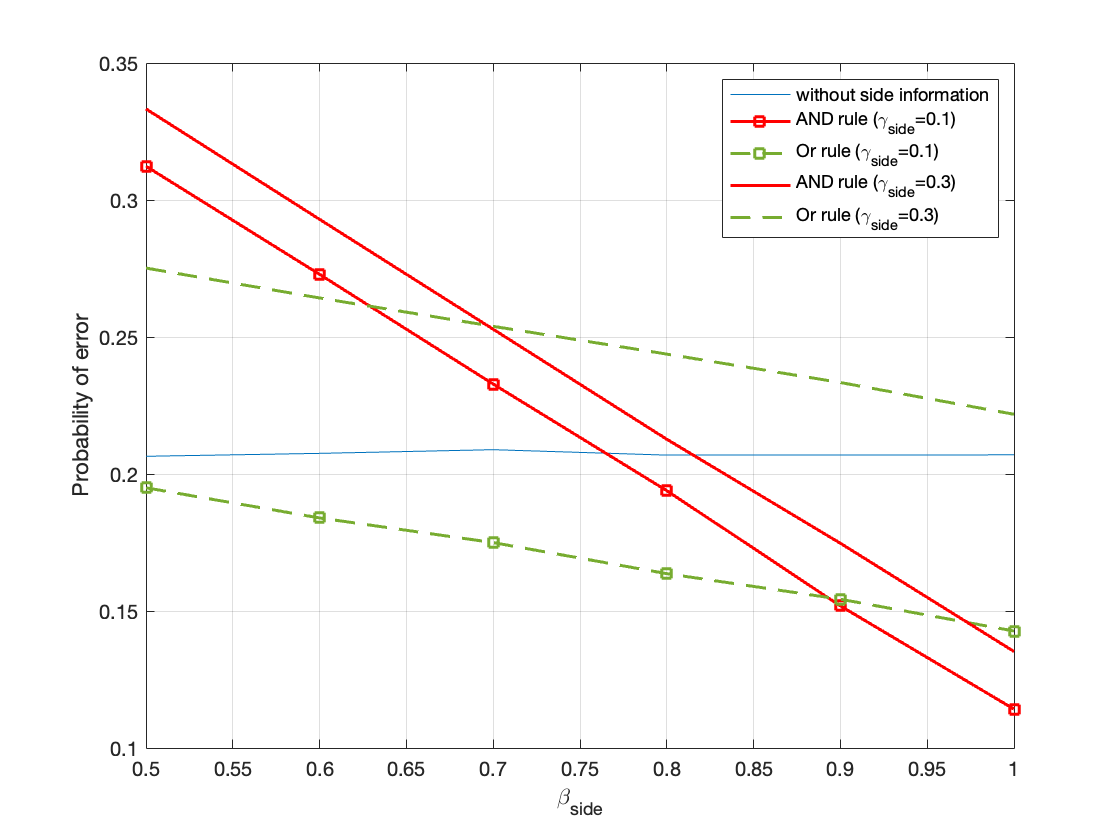}}
\caption{The average probability of error versus $\beta_{side}$ given different values of $\gamma_{side}$ for any human sensor $m$ without side information, as well as for any human sensor $m$ that uses OR or AND operations.}
\label{fig:7}
\end{figure}

The impact of different operations while incorporating the individual performance of a human is illustrated in Fig. \ref{fig:7}. The relationships among the error probability, the detection probability of the side information $\beta_{side}$, and the false alarm probability of the side information $\gamma_{side}$ are shown. It can be observed that given certain values of $\gamma_{side}$, the error probability of a human sensor decreases as $\beta{side}$ increases for both OR and AND operations. Given certain parameters $(\beta_{side},\gamma_{side})$, we can make a better choice among OR operation, AND operation, and no operation (i.e., no side information is utilized). For example, no operation is a better choice given $\beta_{side}\leq0.81$ and $\gamma_{side}=0.1$ and AND operation is a better choice given $\beta_{side}\geq0.9$ and $\gamma_{side}=0.3$. Our results shown in Fig. \ref{fig:7} are also consistent with Theorem \ref{them:comapre_err} we obtained earlier.

\section{CONCLUSION}
In this paper, we have proposed a belief-updating scheme in a human-machine hierarchical network. The local decisions from physical sensors served as reference decisions to improve the quality of human sensor decisions. At the same time, the belief that each physical sensor is malicious was updated during the decision-making process. The impact of side information from an individual human sensor and comparing different operations used to incorporate the side information were also analyzed. Simulation results showed that the quality of human sensors' decisions could be improved by employing the proposed scheme even when most physical sensors in the system are malicious. Moreover, our proposed scheme did not require the knowledge of the actual fraction of malicious physical sensors to guarantee the performance of our proposed scheme. Hence, the proposed scheme can successfully defend against Byzantine attacks and improve the quality of human sensors' decisions.
\bibliography{refer.bib}
\bibliographystyle{IEEEtran}

\end{document}